\documentclass[twocolumn,conference]{IEEEtran}
\usepackage[T1]{fontenc}
\usepackage{mathtools}
\usepackage{amsmath}
\usepackage{amssymb}
\usepackage{graphicx}
\usepackage{wasysym}
\usepackage[unicode=true,
 bookmarks=true,bookmarksnumbered=true,bookmarksopen=true,bookmarksopenlevel=1,
 breaklinks=false,pdfborder={0 0 0},pdfborderstyle={},backref=false,colorlinks=false]
 {hyperref}
\hypersetup{pdftitle={Your Title},
 pdfauthor={Your Name},
 pdfnewwindow=true, pdfstartview=XYZ, plainpages=false}

\makeatletter

\providecommand{\tabularnewline}{\\}
\newcommand{\lyxdot}{.}

\usepackage[noadjust]{cite}
\usepackage[caption=false]{subfig}

\@ifundefined{showcaptionsetup}{}{%
 \PassOptionsToPackage{caption=false}{subfig}}
\usepackage{subfig}
\makeatother

\begin{document}

\title{Fine-Grained vs. Average Reliability for V2V Communications around
Intersections}

\author{Mouhamed~Abdulla and Henk~Wymeersch\\
Department of Electrical Engineering, Division of Communication and
Antenna Systems\\
Chalmers University of Technology, Gothenburg, Sweden\\
Email: \{mouhamed,henkw\}@chalmers.se}
\maketitle
\begin{abstract}
Intersections are critical areas of the transportation infrastructure
associated with 47\% of all road accidents. Vehicle-to-vehicle (V2V)
communication has the potential of preventing up to 35\% of such serious
road collisions. In fact, under the 5G/LTE Rel.15+ standardization,
V2V is a critical use-case not only for the purpose of enhancing road
safety, but also for enabling traffic efficiency in modern smart cities.
Under this anticipated 5G definition, high reliability of 0.99999
is expected for semi-autonomous vehicles (i.e., driver-in-the-loop).
As a consequence, there is a need to assess the reliability, especially
for accident-prone areas, such as intersections. We unpack traditional
average V2V reliability in order to quantify its related fine-grained
V2V reliability. Contrary to existing work on infinitely large roads,
when we consider finite road segments of significance to practical
real-world deployment, fine-grained reliability exhibits bimodal behavior.
Performance for a certain vehicular traffic scenario is either very
reliable or extremely unreliable, but nowhere in relative proximity
to the average performance.
\end{abstract}

\section{\label{Sec1: Introduction-1}Introduction }

Today's machine-driven vehicles are primarily based on LiDARs, sensors,
radars, cameras, GPS, and 3D digital mapping. Nonetheless, these vehicles
are constrained by line-of-sight, and their practicality is significantly
influenced by weather conditions, such as fog, sunbeams, heavy rain
and snow. Meanwhile, vehicle-to-vehicle (V2V) communication is the
\textit{only} technology that enables autonomous vehicles to \textit{see
around corners }in highly urbanized settings in order to detect the
presence of nearby vehicles and take preemptive action to avoid collisions.
In fact, studies suggest that V2V communications may prevent up to
35\% of serious accidents \cite{EU_V2V_35perc}. Moreover, in harsh
weather conditions, V2V communication offers a more reliable alternative
to sensors capability.

Evidently, for the purpose of enhancing road safety and traffic efficiency,
there are numerous V2V use-cases that require careful investigation
and analysis. According to the U.S. Department of Transportation,
data assessed between 2010-2015 suggest that\textit{ nearly half}
of all vehicular accidents occur at \textit{intersections} \cite{DOT_2015_47.24perc}.
What is more surprising is that the rate of intersection accidents
remained steady over the years, and this is despite a growth in the
number of vehicles on the roads and the continuous evolution of accident
avoidance technology. As a result, carefully investigating the reliability
of V2V communications around intersections and, in particular, blind
urban junctions, where the signal strength diminishes significantly
over very short distances, will enable us to explicitly measure the
feasibility of \textit{seeing} vehicles hindered by high-rise infrastructures
near corners.

For scalability and interoperability, standardization of connected
vehicles is exceptionally critical. Currently, two standards are in
direct competition: (i) IEEE 802.11p, also known as direct short range
communication (DSRC) \cite{IEEE802.11P}; and (ii) cellular-V2X (C-V2X)
communications supported by 4G/LTE Rel.14+ \cite{LTERel.14_Jun2015,LTERel.14_Dec2015}.
The ad hoc DSRC standard is defined and ready for utilization, whereas
the ad hoc/network-based C-V2X is still under development for 5G/LTE
Rel.15+ operation aimed near 2020. Meanwhile, highly autonomous vehicles
will rely on machine learning capability, where extensive driving
experience will be shared wirelessly to the cloud and to other units
using V2X capability at multi-Gbps data rates. Other vehicles connected
to the network will immediately upgrade their transportation system
regarding a particular geographical region of interest without the
need of time-consuming knowledge acquisition. For such intricate network,
the potential of 5G/C-V2X is promising and directly suitable to these
stipulated requirements.

Irrespective of the adapted standard, we ultimately need to develop
analytical expressions that will serve as a useful mechanism to accurately
quantify the extent of reliability for a certain V2V communication
link. This analysis will also aid in identifying the contribution
of relevant parameters for the purpose of designing and reconfiguring
the vehicular ad hoc network (VANET). Road traffic modeling based
on point processes is well suited to study such problems where techniques
from stochastic geometry have been applied (e.g. \cite{Bartek09_VANET,Bartek12_VANET,MoeWin2013,Haenggi2016_DSRC,Erik_GC15,Moe_GC16}).
As for intersections, they were explicitly considered in \cite{Erik_GC15},
though only for suburban/rural scenarios over infinitely long roads.
However, for the analytical expressions to have practical real-world
relevance, they must build on plausible VANET scenarios coupled with
channel models validated by precise measurement campaigns \cite{Tufvesson2011IEEEMag,Tufvesson2011TVT}.

The above-mentioned works in stochastic geometry allow the evaluation
of the \emph{average reliability}, obtained by averaging over different
fading realizations and node placements. This average reliability
may obscure the performance for specific node configuration \cite{Haenggi2016_MetaDistribution},
referred to as the \emph{fine-grained reliability}. In this paper,
we perform a study of the fine-grained reliability, dedicated to urban
intersections which have particular propagation characteristics \cite{Tufvesson_4IntersectionCases_2010,UrbanIntersection_OtherGerman11,BMW2011_Virtual_11p,Tfvesson-Abbas2013_Sweden},
complementing and generalizing the results in \cite{Erik_GC15,Moe_GC16}. 

\section{\label{Sec2: System Model-1}System Model}

\subsection{\label{Sec2.1: Network Model}Network Model}

The VANET formed around the intersection is described as follows.
The position of the transmitter (TX) can be anywhere on the horizontal
or vertical road. Without loss of generality, the receiver (RX) is
confined to the horizontal road. Thus, $\mathbf{x}_{\text{\ensuremath{\mathrm{tx}}}}\!=\!x_{\mathrm{tx}}\mathbf{e}_{\mathrm{x}}+y_{\mathrm{tx}}\mathbf{e}_{\mathrm{y}}$
and $\mathbf{x}_{\mathrm{rx}}\!=\!x_{\mathrm{rx}}\mathbf{e}_{\mathrm{x}}$,
where $x_{\mathrm{tx}},x_{\mathrm{rx}},y_{\mathrm{tx}}\in\mathbb{R}$,
such that $x_{\mathrm{tx}}y_{\mathrm{tx}}\!=\!0$, where $\mathbf{e}_{\mathrm{x}}\!=\![1\,0]^{\mathrm{T}}$,
$\mathbf{e}_{\mathrm{y}}\!=\![0\,1]^{\mathrm{T}}$. Other traffic
vehicles are randomly positioned on both horizontal and vertical roads
and follow a homogeneous Poisson point process (H-PPP) over bounded
sets $\mathcal{B}_{\mathrm{x}}=\left\{ x\in\mathbb{R},\,R_{\mathrm{x}}\in\mathbb{R}^{+}\bigl|\left|x\right|\leq R_{\mathrm{x}}\right\} $
and $\mathcal{B}_{\mathrm{y}}=\left\{ y\in\mathbb{R},\,R_{\mathrm{y}}\in\mathbb{R}^{+}\bigl|\left|y\right|\leq R_{\mathrm{y}}\right\} $,
such that $R_{\mathrm{x}}$ and $R_{\mathrm{y}}$ are road segments
of the intersection region, and the vehicular traffic intensities
are respectively given by $\lambda_{\mathrm{x}}$ and $\lambda_{\mathrm{y}}$.
The interfering vehicles follow an Aloha MAC protocol\footnote{Resource selection for DSRC is based on CSMA with collision avoidance;
and C-V2X defined by 3GPP-PC5 interface relies on semi-persistent
transmission with relative energy-based selection. Nonetheless, for
the purpose of preliminary analysis, we only consider an Aloha MAC
protocol.} and can transmit independently with a probability $p_{\mathrm{I}}\!\in\!\left[0,1\right]$.
The following shorthand notations are accordingly used to refer to
the geometry of interfering vehicles on each road, modeled by thinned
H-PPPs
\begin{align}
\Phi_{\mathrm{x}} & =\left\{ \mathbf{x}_{i}\right\} _{i=1,2,\ldots,n}\ \in\mathbb{R}^{n}\:\sim\thinspace\textrm{PPP}\left(p_{\mathrm{I}}\lambda_{\mathrm{x}},\mathcal{B}_{\mathrm{x}}\right)\label{eq: PPP-X}\\
\Phi_{\mathrm{y}} & =\left\{ \mathbf{x}_{j}\right\} _{j=1,2,\ldots,m}\in\mathbb{R}^{m}\sim\thinspace\textrm{PPP}\left(p_{\mathrm{I}}\lambda_{\mathrm{y}},\mathcal{B}_{\mathrm{y}}\right),\label{eq: PPP-Y}
\end{align}
such that $n$ and $m$ are random Poisson distributed integers with
mean $p_{\mathrm{I}}\lambda\left|\mathcal{B}\right|$, where $\left|\mathcal{B}\right|$
is the Lebesgue measure of bounded set $\mathcal{B}$. All vehicles,
including TX, broadcast with the same power level $P_{\circ}$. The
signal-to-interference-plus-noise-ratio ($\mathsf{SINR}$) threshold
for reliable detection at the RX is set to $\beta$, in the presence
of additive white Gaussian noise (AWGN) with power $N_{\circ}$. The
$\mathsf{SINR}$ depends on the propagation channel, which we describe
in the next subsection.

\subsection{\label{Sec2.2: Channel Models}Channel Models}

The detected power at the RX from an active TX located at $\mathbf{x}$
is modeled by $P_{\mathrm{rx}}\left(\mathbf{x},\mathbf{x}_{\mathrm{rx}}\right)=P_{\circ}\ell_{\mathrm{ch}}\left(\mathbf{x},\mathbf{x}_{\mathrm{rx}}\right)$,
which depends on transmit power $P_{\circ}$ and channel losses $\ell_{\mathrm{ch}}\left(\mathbf{x},\mathbf{x}_{\mathrm{rx}}\right)$.
The components of channel losses are: average path loss $\ell_{\mathrm{pl}}\left(\mathbf{x},\mathbf{x}_{\mathrm{rx}}\right)$
that captures signal attenuation, shadow fading $\ell_{\mathrm{s}}\left(\mathbf{x},\mathbf{x}_{\mathrm{rx}}\right)$
that captures the impact of random obstacles, and small-scale fading
$\ell_{\mathrm{f}}\left(\mathbf{x}\right)$ that accounts for the
non-coherent addition of signal components. For the purpose of tractability,
we implicitly consider shadow fading to be inherent within the H-PPP,
and thus regard $\ell_{\mathrm{ch}}\left(\mathbf{x},\mathbf{x}_{\mathrm{rx}}\right)\simeq\ell_{\mathrm{pl}}\left(\mathbf{x},\mathbf{x}_{\mathrm{rx}}\right)\ell_{\mathrm{f}}\left(\mathbf{x}\right)$
\cite{RossEquivalence17}. We model $\ell_{\mathrm{f}}\left(\mathbf{x}\right)\sim\textrm{Exp}\left(1\right)$
as Rayleigh fading, independent with respect to $\mathbf{x}$. The
path loss $\ell_{\mathrm{pl}}\left(\mathbf{x},\mathbf{x}_{\mathrm{rx}}\right)$
for different channel environments is described below.

\subsubsection{Suburban/Rural Channel}

The average path loss model for V2V propagation adheres to an inverse
power-law \cite{Tufvesson2011TVT,Tufvesson2011IEEEMag}; thus
\begin{align}
\ell_{\mathrm{pl}}^{\mathrm{s}}\left(\mathbf{x},\mathbf{x}_{\mathrm{rx}}\right)=A_{\circ}\left\Vert \mathbf{x_{\mathrm{rx}}}-\mathbf{x}\right\Vert ^{-\alpha} & \ \ \ \ \mathbf{x}\neq\mathbf{x_{\mathrm{rx}}}\mathrm{.}\label{eq: Channel - Suburban}
\end{align}
In this expression, $\left\Vert \cdot\right\Vert $ is the $l_{2}$-norm,
$A_{\circ}$ corresponds to the LOS/WLOS path loss coefficient, which
is primarily a function of operating frequency $f_{\circ}$, path
loss exponent $\alpha\!>\!1$, reference distance $d_{\circ}$, and
antenna heights $h\left(\mathbf{x}\right)$ and $h\left(\mathbf{x}_{\mathrm{rx}}\right)$.

\subsubsection{Urban Channel}

For metropolitan intersections where the concentration of high-rise
and impenetrable metallic-based buildings and structures are prevalent,
the previous Euclidean model is rather unrealistic. Real-world measurements
of V2V communications operating at 5.9\,GHz were conducted at different
urban intersection locations, leading to the VirtualSource11p path
loss model \cite{BMW2011_Virtual_11p,Tfvesson-Abbas2013_Sweden},
adapted here\footnote{\noindent The model in (\ref{eq: Urban PL model}) exhibits discontinuities.
A mixture (linear weighting) of these models can be used to avoid
these discontinuities, though this is not considered in this paper.} as
\begin{align}
 & \ell_{\mathrm{pl}}^{\mathrm{u}}\left(\mathbf{x},\mathbf{x}_{\mathrm{rx}}\right)=\nonumber \\
 & \ \ \ \ \ \ \ \begin{cases}
A_{\circ}^{\prime}\left(\left\Vert \mathbf{x}\right\Vert \cdot\left\Vert \mathbf{x}_{\mathrm{rx}}\right\Vert \right)^{-\alpha} & \min\left(\left\Vert \mathbf{x}\right\Vert =\left|y\right|,\left\Vert \mathbf{x}_{\mathrm{rx}}\right\Vert \right)>\triangle\\
A_{\circ}\left(\left\Vert \mathbf{x}\right\Vert +\left\Vert \mathbf{x}_{\mathrm{rx}}\right\Vert \right)^{-\alpha} & \min\left(\left\Vert \mathbf{x}\right\Vert =\left|y\right|,\left\Vert \mathbf{x}_{\mathrm{rx}}\right\Vert \right)\leq\triangle\ \ \ \\
A_{\circ}\left\Vert \mathbf{x}_{\mathrm{rx}}-\mathbf{x}\right\Vert ^{-\alpha} & \left\Vert \mathbf{x}\right\Vert =\left|x\right|;\ \mathbf{x}\neq\mathbf{x}_{\mathrm{rx}},
\end{cases}\label{eq: Urban PL model}
\end{align}
where $\triangle\le\min\left(R_{\mathrm{x}},R_{\mathrm{y}}\right)$
is the break-point distance (typically on the order of the road width),
and $A_{\circ},A'_{\circ}$ are suitable path loss coefficients \cite{Moe_arXiv2017}.
The first case in (\ref{eq: Urban PL model}) refers to NLOS link,
the second to WLOS, and the third to LOS.

\noindent 

\section{Transmission Reliability}

\subsection{\label{Sec3.2: Expected Reliability}Average Reliability}

The goal is to determine the success probability $\mathcal{P}_{\mathrm{c}}\left(\beta,\mathbf{x}_{\text{\ensuremath{\mathrm{tx}}}},\mathbf{x}_{\mathrm{rx}}\right)\triangleq\Pr\left(\mathsf{SINR}\geq\beta\right)$,
where $\Pr\left(\cdot\right)$ is averaged over small-scale fading
of the channel $\ell_{\mathrm{f}}$ and point processes $\Phi_{\mathrm{x}}$
and $\Phi_{\mathrm{y}}$; and where
\begin{align}
\mathsf{SINR} & =\frac{\ell_{\mathrm{f}}\left(\mathbf{x}_{\text{\ensuremath{\mathrm{tx}}}}\right)\ell_{\mathrm{pl}}\left(\mathbf{x}_{\text{\ensuremath{\mathrm{tx}}}},\mathbf{x}_{\mathrm{rx}}\right)}{{\displaystyle \sum_{\mathbf{x}\in\Phi_{\mathrm{x}}\cup\Phi_{\mathrm{y}}}\ell_{\mathrm{f}}\left(\mathbf{x}\right)\ell_{\mathrm{pl}}\left(\mathbf{x},\mathbf{x}_{\mathrm{rx}}\right)+\gamma_{\circ}}},\label{eq: SINR =00003D S/(I+N)}
\end{align}
in which $\gamma_{\circ}\!\!=\!\!N_{\circ}/P_{\circ}$. We introduce
a normalized aggregate interference associated with a process $\Phi$
as: $I\left(\Phi\right)\triangleq\sum_{\mathbf{x}\in\Phi}\ell_{\mathrm{f}}\left(\mathbf{x}\right)\ell_{\mathrm{pl}}\left(\mathbf{x},\mathbf{x}_{\mathrm{rx}}\right)$.
Solving for the exponentially distributed fading of the wanted link,
and taking the expectation of the probability with respect to interference,
we get
\begin{align}
\mathcal{P}_{\mathrm{c}}\left(\beta,\mathbf{x}_{\text{\ensuremath{\mathrm{tx}}}},\mathbf{x}_{\mathrm{rx}}\right) & =\exp\bigl(-\beta^{\prime}\gamma_{\circ}\bigr)\thinspace\times\thinspace\mathbb{E}_{I_{\mathrm{x}}}\left\{ \exp\bigl(-\beta^{\prime}I\left(\Phi_{\mathrm{x}}\right)\bigr)\right\} \ \ \ \nonumber \\
 & \times\thinspace\mathbb{E}_{I_{\mathrm{y}}}\left\{ \exp\bigl(-\beta^{\prime}I\left(\Phi_{\mathrm{y}}\right)\bigr)\right\} ,\label{eq: Success Probability: Pc =00003D Po * Px * Py-1}
\end{align}
where $\beta^{\prime}=\beta/\ell_{\mathrm{pl}}\left(\mathbf{x}_{\text{\ensuremath{\mathrm{tx}}}},\mathbf{x}_{\mathrm{rx}}\right)$,
the success probability in the absence of interference is simply $\exp\bigl(-\beta^{\prime}\gamma_{\circ}\bigr)$,
and the remaining factors capture the degradation of the average reliability
due to independent aggregate interference from the horizontal and
vertical roads \cite{Moe_GC16}. 

\subsection{\label{Sec3.3: Granular Reliability}Fine-Grained Reliability}

The success probability $\mathcal{P}_{\mathrm{c}}\left(\beta,\mathbf{x}_{\text{\ensuremath{\mathrm{tx}}}},\mathbf{x}_{\mathrm{rx}}\right)$
provides a high-level performance assessment averaged over all possible
vehicular traffic realizations and channel. Unpacking this average
reliability metric and looking at it at the fine-grained or meta-level
demands that we study the success probability of the individual links.
In other words, we need to explore the \textit{meta distribution }of
the $\mathsf{SINR}$ \cite{Haenggi2016_MetaDistribution}; defined,
using the Palm probability of the point process $\mathbb{P}^{o}\left(\cdot\right)$,
as 
\begin{align}
F_{\mathrm{r}}\left(\beta,p\right) & \thinspace\triangleq\thinspace\mathbb{P}^{o}\Bigl(\thinspace\Pr\left(\mathsf{SINR}\geq\beta\thinspace|\thinspace\Phi_{\mathrm{x}}\cup\Phi_{\mathrm{y}}\right)\geq p\thinspace\Bigr)\mathrm{.}\label{eq: Meta Distribution (Success Probability)}
\end{align}
Introducing $p_{\mathrm{c}}\left(\beta\right)\triangleq\Pr\left(\mathsf{SINR}\geq\beta\thinspace|\thinspace\Phi_{\mathrm{x}}\cup\Phi_{\mathrm{y}}\right)$
and $p_{\mathrm{out}}\left(\beta\right)=1-p_{\mathrm{c}}\left(\beta\right)$
respectively as the conditional success probability and conditional
outage probability, given the point process $\Phi_{\mathrm{x}}\cup\Phi_{\mathrm{y}}$,
and where $p\!\in\!\left[0,1\right]$ is a conditional success reliability
constraint. Thus, the meta distribution $F_{\mathrm{r}}\left(\beta,p\right)$
is basically the fraction of vehicular traffic realizations that achieve
reliability, in which reliability is prescribed by the target value
assigned to $p$, and $\beta$ refers to the $\mathsf{SINR}$ threshold.

Using the Palm expectation $\mathbb{E}^{o}\left(\cdot\right)$, we
can obtain the average reliability of success from the meta distribution.
In other words, the first moment of $p_{\mathrm{c}}\left(\beta\right)$
is the average reliability of success, i.e., $\mathbb{E}^{o}\left(p_{\mathrm{c}}\left(\beta\right)\right)=\mathcal{P}_{\mathrm{c}}\left(\beta,\mathbf{x}_{\text{\ensuremath{\mathrm{tx}}}},\mathbf{x}_{\mathrm{rx}}\right)$.
Meanwhile, rather than obtaining the exact meta distribution through
the Gil-Pelaez theorem \cite{GilPelaez1951}, recent meta distribution
related works suggest that the moments could be used to approximate
$F_{\mathrm{r}}\left(\beta,p\right)$. In fact, the Beta probability
density function (which only requires the first and second moments)
is reported to yield high accuracy \cite{Haenggi2016_MetaDistribution,Haenggi2017_MSalihi,Haenggi2017_MDWang}:
\begin{equation}
F_{\mathrm{r}}\left(\beta,p\right)\thinspace\approx\thinspace\frac{p^{a-1}(1-p)^{b-1}}{B(a,b)},
\end{equation}
where $B(a,b)$ is the Beta function, which serves as a normalization
constant. The parameters of the Beta distribution can be estimated
from $N$ realizations of $x\coloneqq\Pr\left(\mathsf{SINR}\geq\beta\thinspace|\thinspace\Phi_{\mathrm{x}}\cup\Phi_{\mathrm{y}}\right)$,
say $x_{1},\ldots,x_{N}\in[0,1]$ using a methods of moments estimator.
Introducing the estimates of the mean $\bar{x}=\sum_{n}x_{n}/N$,
variance $v=\sum_{n}(x_{n}-\bar{x})^{2}/(N-1)$, and odds $o=(1-\bar{x})/\bar{x}$
, we find that $\hat{a}(\mathbf{x})=\bar{x}\left(o/v-1\right)$ and
$\hat{b}(\mathbf{x})=o\,\hat{a}(\mathbf{x})$, provided the estimates
are non-negative. 

\begin{table}
\caption{Simulation Parameters\label{Tab: Simulation Parameters}}

\resizebox{0.5\textwidth}{!}{%
\begin{centering}
{\scriptsize{}}%
\begin{tabular}{ll}
\hline 
\multicolumn{2}{l}{\textbf{\scriptsize{}system parameters}}\tabularnewline
\hline 
\hline 
{\scriptsize{}target success probability} & {\scriptsize{}$\mathcal{P}_{\mathrm{target}}=0.9$}\tabularnewline
{\scriptsize{}transmit power} & {\scriptsize{}$P_{\circ}=20$\,dBmW}\tabularnewline
{\scriptsize{}AWGN floor} & {\scriptsize{}$N_{\circ}=-99$\,dBmW}\tabularnewline
{\scriptsize{}RX sensitivity} & {\scriptsize{}$\beta=8$\,dB (if $B=40$\,MHz; $r_{\mathrm{th}}\simeq115$\,Mbps)}\tabularnewline
\hline 
\hline 
\multicolumn{2}{l}{\textbf{\scriptsize{}channel propagation}}\tabularnewline
\hline 
\hline 
{\scriptsize{}operating frequency} & {\scriptsize{}$f_{\circ}=5.9$\,GHz}\tabularnewline
{\scriptsize{}reference distance} & {\scriptsize{}$d_{\circ}=10$\,m}\tabularnewline
{\scriptsize{}break-point distance} & {\scriptsize{}$\triangle=15$\,m}\tabularnewline
{\scriptsize{}path loss exponent} & {\scriptsize{}$\alpha=2$\,(suburban); $1.68$\,(urban)}\tabularnewline
{\scriptsize{}LOS/WLOS path loss coefficient} & {\scriptsize{}$A_{\circ}\!=\!-37.86+10\alpha$\,\,dBm}\tabularnewline
{\scriptsize{}NLOS path loss coefficient} & {\scriptsize{}$A_{\circ}^{\prime}\!=\!-38.32+\left(7\!+\!10\log_{10}\triangle\right)\!\alpha$\,\,dBm}\tabularnewline
\hline 
\hline 
\multicolumn{2}{l}{\textbf{\scriptsize{}vehicular traffic and geometry}}\tabularnewline
\hline 
\hline 
{\scriptsize{}traffic intensity} & {\scriptsize{}$\lambda=0.01$\,\#\,/\,m}\tabularnewline
{\scriptsize{}size of road segment} & {\scriptsize{}$R=200$\,m\,(practical); $10$\,km\,(stress-test)}\tabularnewline
{\scriptsize{}RX distance from junction point} & {\scriptsize{}$\left\Vert \mathbf{x}_{\mathrm{rx}}\right\Vert =50$\,m}\tabularnewline
{\scriptsize{}max. separation for reliable V2V com.} & {\scriptsize{}$d_{\mathrm{target}}=100$\,m}\tabularnewline
{\scriptsize{}max. TX/RX Manhattan separation} & {\scriptsize{}$d_{\mathrm{max}}=140$\,m}\tabularnewline
\hline 
\end{tabular}
\par\end{centering}{\scriptsize \par}
}
\end{table}

\section{\label{Sec6: Simulations and Discussion}Simulations and Discussion}

\subsection{Simulation Setup}

Using the parameters shown in Table \ref{Tab: Simulation Parameters},
we evaluate the success probability under various conditions and scenarios.
In particular, we set the vehicular traffic on both roads to be the
same, i.e., $\lambda_{\mathrm{x}}\!=\!\lambda_{\mathrm{y}}\!=\!\lambda\!=\!0.01$\,\#/m.
For identical road segments $R_{\mathrm{x}}\!=\!R_{\mathrm{y}}\!=\!R$,
we consider $R\!\in\!\{200\,\mathrm{m},10\,\mathrm{km}\}$. Next,
we assume a fixed RX on the horizontal road, with $\mathbf{x}_{\mathrm{rx}}\!=\![-50,0]^{\mathrm{T}}$\,m;
and a TX that could take different positions, up to a Manhattan separation
of $d_{\mathrm{max}}\!=\!140$\,m away from the RX, i.e., starting
at $\mathbf{x}_{\mathrm{tx}}\!=\![-50,0]^{\mathrm{T}}$\,m via $\mathbf{x}_{\mathrm{tx}}\!=\![0,0]^{\mathrm{T}}$~m
up to $\mathbf{x}_{\mathrm{tx}}\!=\![0,+90]^{\mathrm{T}}$~m. To
ensure a tolerable worst-case level of average performance, the success
probability must achieve a certain preassigned target value, here
set to 
\begin{equation}
\mathcal{P}_{\mathrm{c}}\left(\beta,\mathbf{x}_{\text{\ensuremath{\mathrm{tx}}}},\mathbf{x}_{\mathrm{rx}}\right)\thinspace\geq\thinspace\mathcal{P}_{\mathrm{target}}=0.9,
\end{equation}
over the intersection deployment region specified by $\mathcal{B}_{\mathrm{x}}\cup\mathcal{B}_{\mathrm{y}}$,
and for all V2V communication pairs under consideration with positions
$\mathbf{x}_{\mathrm{tx}}$ and $\mathbf{x}_{\mathrm{rx}}$. As design
parameters, we consider the transmit probability $p_{\mathrm{I}}$
and its relation to road segments $R_{\mathrm{x}}$ and $R_{\mathrm{y}}$.
Solving for the transmit probability in this design criteria, we find
that $p_{\mathrm{I}}\leq p_{\mathrm{I}}^{\ast}\left(R\right)$, where
the optimum probability was derived in \cite{Moe_arXiv2017}. In other
words, with $p_{\mathrm{I}}\leq p_{\mathrm{I}}^{\ast}\left(R\right)$,
$\mathcal{P}_{\mathrm{c}}\left(\beta,\mathbf{x}_{\text{\ensuremath{\mathrm{tx}}}},\mathbf{x}_{\mathrm{rx}}\right)\geq0.9$,
provided the RX remains fixed, the TX can have different positions,
while $\left\Vert \mathbf{x}_{\mathrm{rx}}-\mathbf{x}_{\mathrm{tx}}\right\Vert _{1}\leq\left\Vert \mathbf{x}_{\mathrm{rx}}-\tilde{\mathbf{x}}_{\mathrm{tx}}\right\Vert _{1}$,
where $\left\Vert \cdot\right\Vert _{1}$ is the $l_{1}$-norm and
$\tilde{\mathbf{x}}_{\mathrm{tx}}$ is the TX position at target (i.e.,
at the worst-case position where target reliability is still fulfilled).

\subsection{Sensitivity of Fine-Grained Reliability to TX/RX Separation}

We aim to gain insight into the meta distribution for the $\mathbf{x}_{\text{\ensuremath{\mathrm{tx}}}}$
and $\mathbf{x}_{\mathrm{rx}}$ position variables, obtained from
outage probability conditioned on a traffic realization, and as a
function of TX/RX Manhattan separation. For every TX/RX position pair,
we consider 10000 PPPs and for each PPP, 5000 fading realizations.
The outage probability per intersection traffic of the point process,
i.e., $\Pr\left(\mathsf{SINR}\leq\beta\thinspace|\thinspace\Phi_{\mathrm{x}}\cup\Phi_{\mathrm{y}}\right)$,
can therefore be determined from extensive Monte Carlo simulations.

\begin{figure*}
\begin{centering}
\subfloat[practical deployment: $R=200$ m.\label{fig6: Granular Reliability--200-1}]{\raggedright{}\includegraphics[width=1\columnwidth]{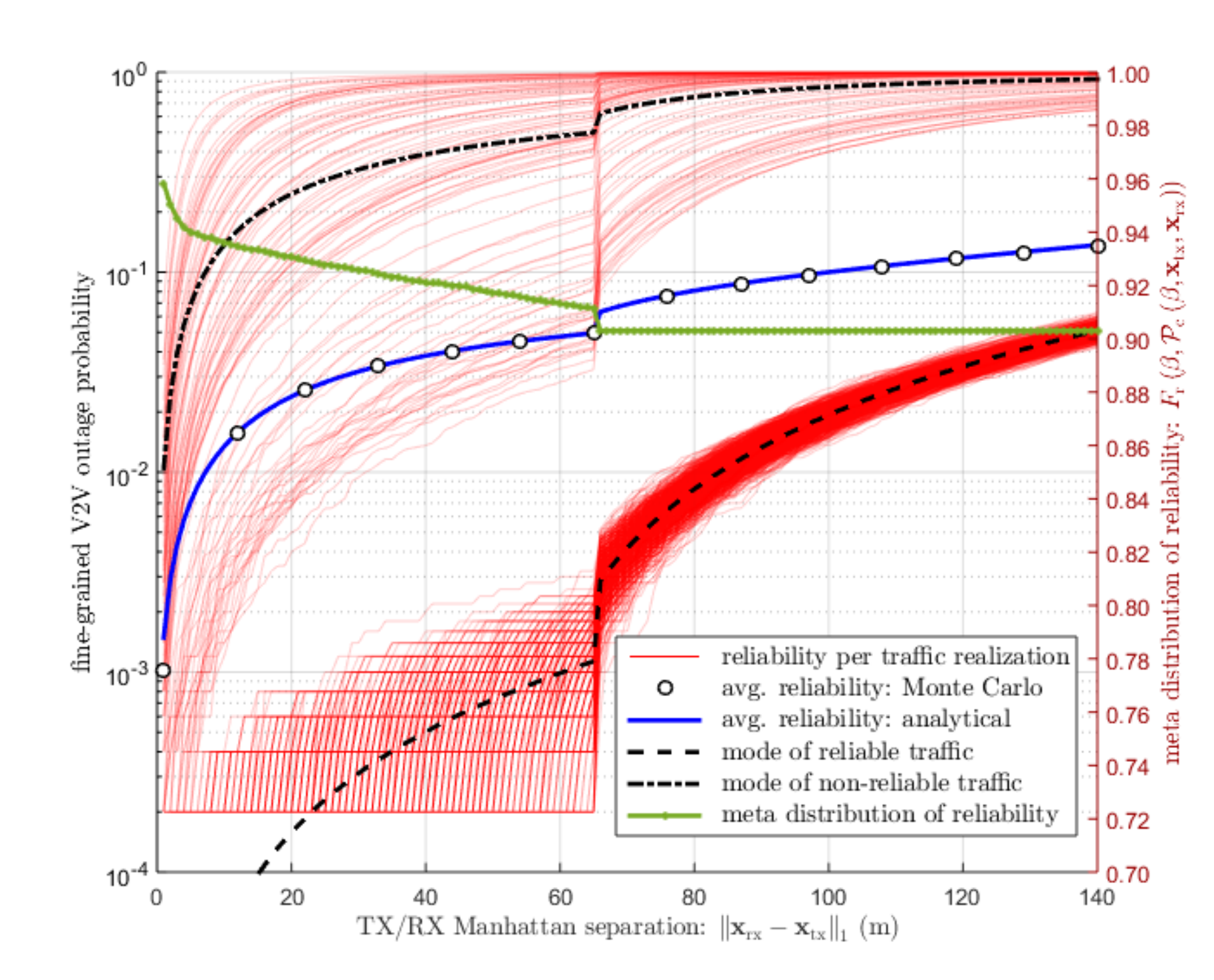}}~~~~~~~\subfloat[stress-test: $R=10$ km.\label{fig6: Granular Reliability--10km}]{\raggedright{}\includegraphics[width=1\columnwidth]{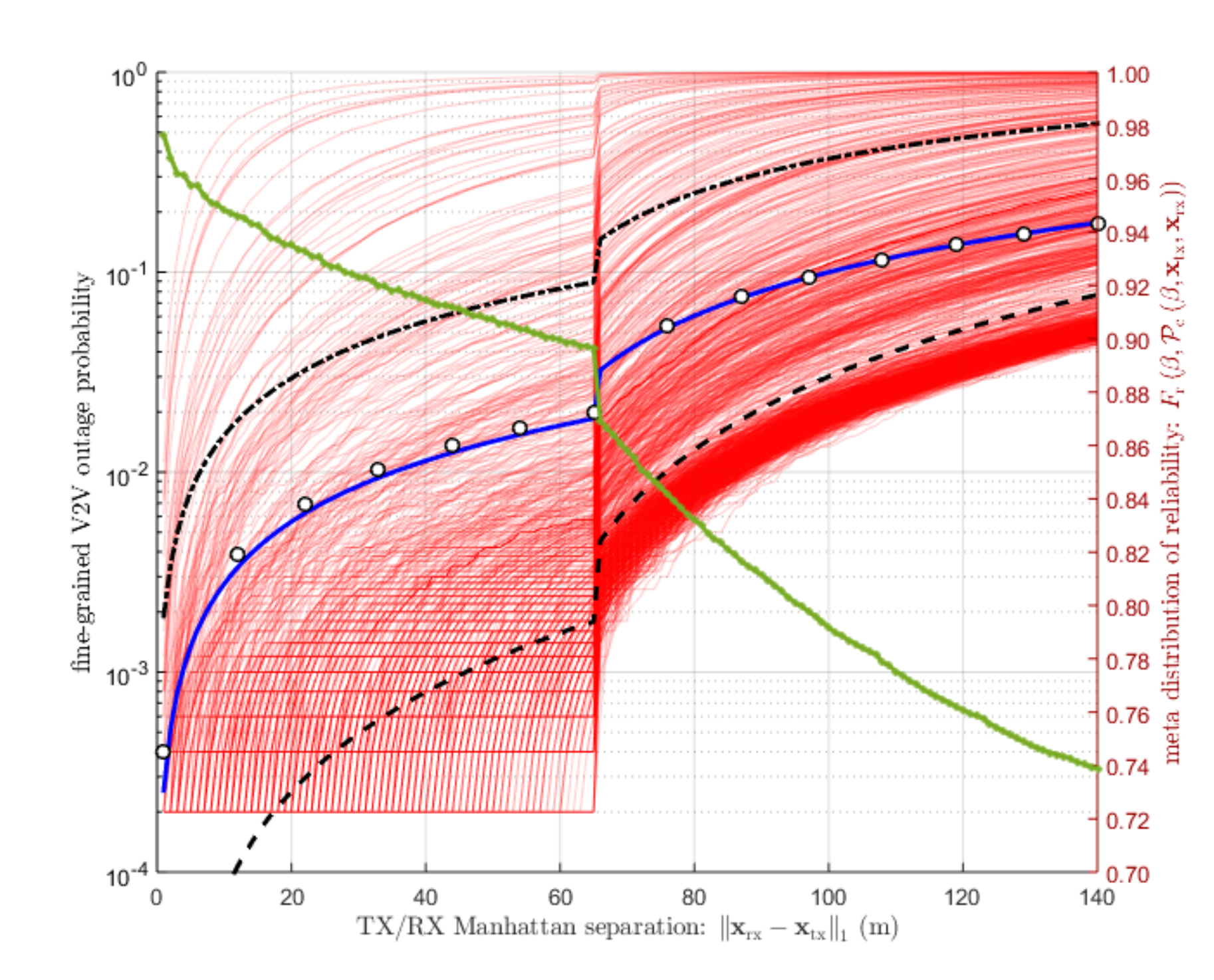}}
\par\end{centering}
\centering{}\caption{Fine-grained reliability under network design as a function of TX/RX
separation for urban intersection. The meta distribution of reliability
is also shown over different V2V separations.}
\end{figure*}

In Fig.~\ref{fig6: Granular Reliability--200-1}, considering the
urban case and $R=200$\,m (with $p_{\mathrm{I}}^{\ast}\left(R\right)=0.013$
to ensure an average success probability of 0.9 at $\left\Vert \mathbf{x}_{\mathrm{rx}}-\tilde{\mathbf{x}}_{\mathrm{tx}}\right\Vert _{1}=d_{\mathrm{target}}=100\,\mathrm{m}$),
we plot the fine-grained outage probability (with an axis on the left),
and the meta distribution (with an axis on the right) both as a function
of $\left\Vert \mathbf{x}_{\mathrm{rx}}-\mathbf{x}_{\mathrm{tx}}\right\Vert _{1}$.
We observe that the average outage probability $\mathcal{P}_{\mathrm{c}}\left(\beta,\mathbf{x}_{\text{\ensuremath{\mathrm{tx}}}},\mathbf{x}_{\mathrm{rx}}\right)$
(in blue) increases with $\left\Vert \mathbf{x}_{\mathrm{rx}}-\mathbf{x}_{\mathrm{tx}}\right\Vert _{1}$
and reaches the target value of 0.1 when $\left\Vert \mathbf{x}_{\mathrm{rx}}-\mathbf{x}_{\mathrm{tx}}\right\Vert _{1}=d_{\mathrm{target}}$.
The corresponding meta distribution $F_{\mathrm{r}}\left(\beta,\mathcal{P}_{\mathrm{c}}\left(\beta,\mathbf{x}_{\text{\ensuremath{\mathrm{tx}}}},\mathbf{x}_{\mathrm{rx}}\right)\right)$
(in green) decreases, indicating that for low $\left\Vert \mathbf{x}_{\mathrm{rx}}-\mathbf{x}_{\mathrm{tx}}\right\Vert _{1}$,
more PPPs achieve an outage below the average than at high separation
values of $\left\Vert \mathbf{x}_{\mathrm{rx}}-\mathbf{x}_{\mathrm{tx}}\right\Vert _{1}$.
Meanwhile, the sharp transitions of both curves at 65\,m is due to
the crossing of the vehicle from the WLOS region to the blind NLOS
region. The figures also show the realizations of $\Pr\left(\mathsf{SINR}\leq\beta\thinspace|\thinspace\Phi_{\mathrm{x}}\cup\Phi_{\mathrm{y}}\right)$
for 1000 randomly chosen PPPs from the overall total of 10000, and
the process is repeated with intervals of 1\,m over different values
of TX/RX Manhattan separation. From the plots, we see a clear bi-modal
behavior, a PPP traffic realization either leads to communication
reliability: \textit{far better} than the requirement or \textit{far
worse} than the requirement, but not close to the average curve. By
design, the average outage probability is mainly determined by the
fraction of PPPs that meet the performance target, i.e., $F_{\mathrm{r}}\left(\beta,\mathcal{P}_{\mathrm{c}}\left(\beta,\mathbf{x}_{\text{\ensuremath{\mathrm{tx}}}},\mathbf{x}_{\mathrm{rx}}\right)\right)$,
which for $\left\Vert \mathbf{x}_{\mathrm{rx}}-\mathbf{x}_{\mathrm{tx}}\right\Vert _{1}=100\,\mathrm{m}$
evaluates to $F_{\mathrm{r}}\left(\beta,0.9\right)\simeq0.9$. 

In Fig.~\ref{fig6: Granular Reliability--10km}, we show the results
for the stress-test scenario over a very large road segment corresponding
to $R=10$ km (with $p_{\mathrm{I}}^{\ast}\left(R\right)=0.0021$
to ensure an average success probability of 0.9 at $\left\Vert \mathbf{x}_{\mathrm{rx}}-\tilde{\mathbf{x}}_{\mathrm{tx}}\right\Vert _{1}=d_{\mathrm{target}}=100\,\mathrm{m}$).
We note a steeper decline in the meta distribution $F_{\mathrm{r}}\left(\beta,\mathcal{P}_{\mathrm{c}}\left(\beta,\mathbf{x}_{\text{\ensuremath{\mathrm{tx}}}},\mathbf{x}_{\mathrm{rx}}\right)\right)$,
indicating a sharp deterioration of fine-grained reliability due to
a larger interference region. This is also reflected in the outage
probability curves per PPP realization, showing more spread than for
the case of $R=200$\,m. 

Overall, we find that the generally accepted perception that performance
for a particular traffic realization will be in relative proximity
to the average reliability is misleading. This is especially true
for interference limited over a short road segment of relevance to
real-world V2V communications deployment. 

\begin{figure}
\centering{}\includegraphics[width=1.05\columnwidth]{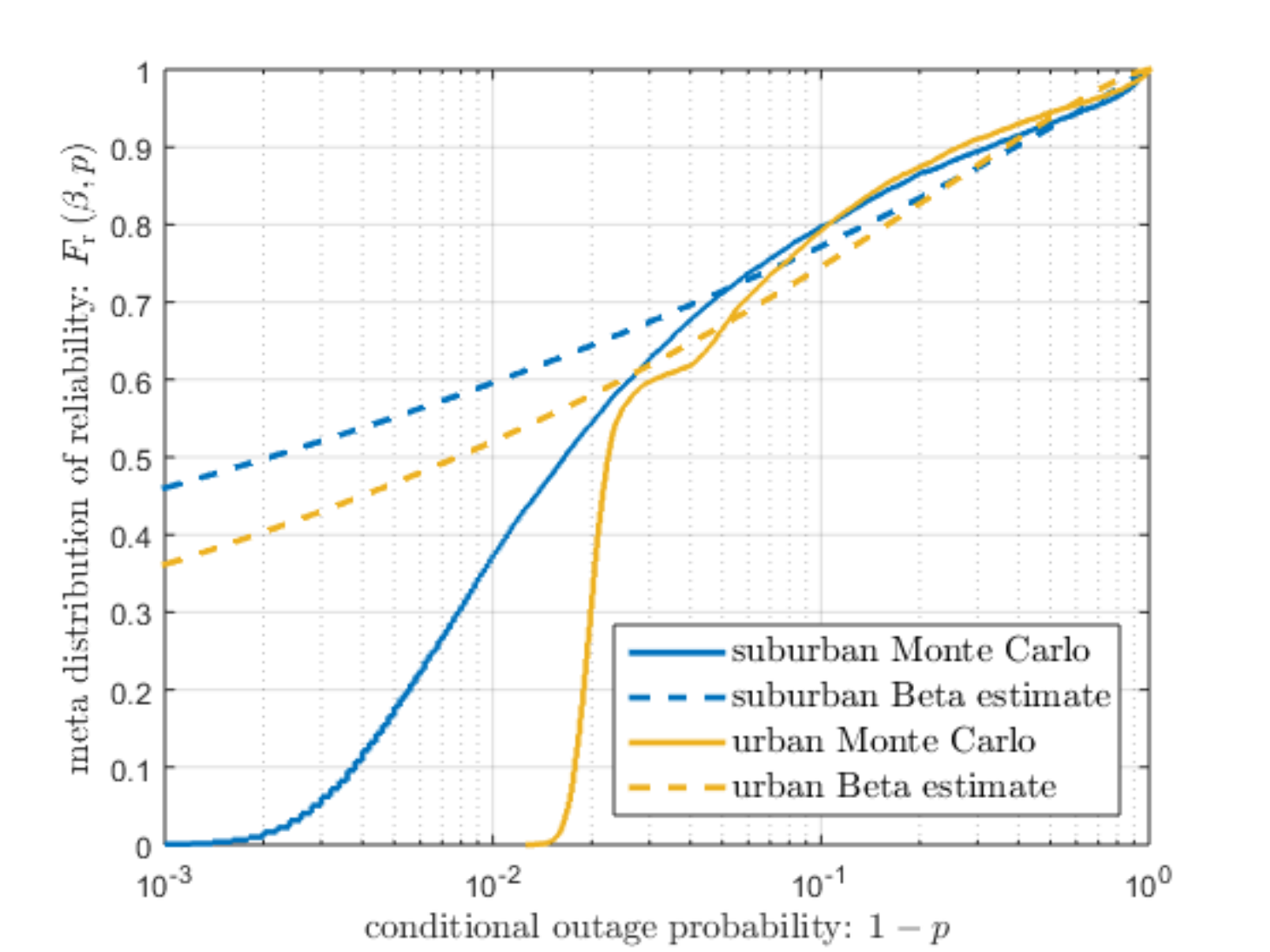}\caption{Meta distribution and Beta distribution approximation of reliability
as a function of conditional outage probability for $R=10$\,km over
urban and suburban intersections.  \label{fig5: MDLong} }
\end{figure}

\subsection{Meta Distribution}

\begin{figure}
\begin{centering}
\includegraphics[width=1.05\columnwidth]{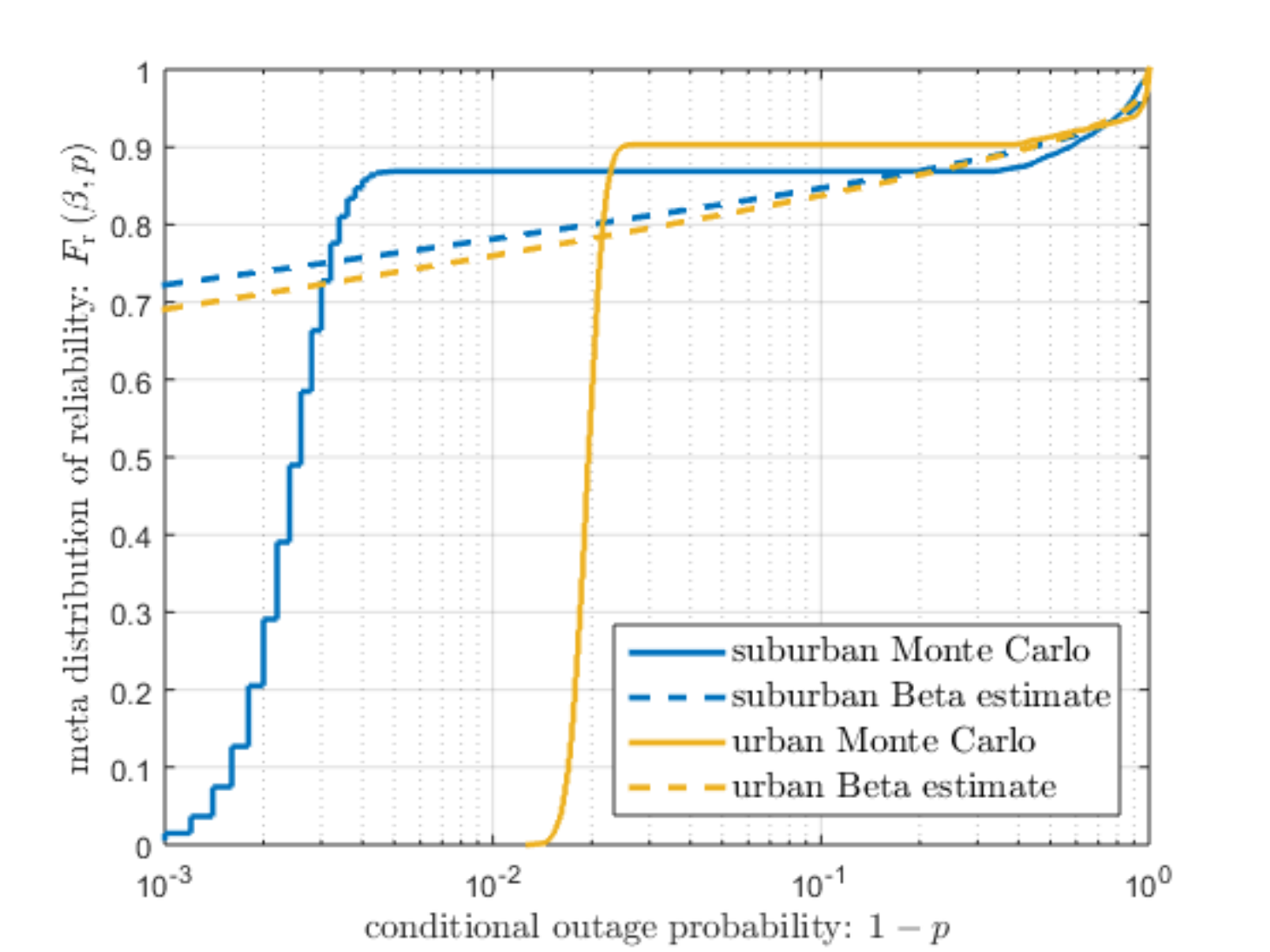}
\par\end{centering}
\vspace{-0.2cm}
\begin{centering}
\caption{Meta distribution and Beta distribution approximation of reliability
as a function of conditional outage probability for $R=200$\,m over
urban and suburban intersections. \label{fig5: MDShort} }
\par\end{centering}
\vspace{-0.3cm}
\end{figure}

We now fix $\left\Vert \mathbf{x}_{\mathrm{rx}}-\mathbf{x}_{\mathrm{tx}}\right\Vert _{1}=100\,\mathrm{m}$
and determine the complete meta distribution, $F_{\mathrm{r}}\left(\beta,p\right)$
from the simulation data. For all cases, the average outage probability
is set to 0.1. We also determine the parameters of the Beta distribution
through moment matching. For $R=10$\,km (see Fig.~\ref{fig5: MDLong}),
we observe that $F_{\mathrm{r}}\left(\beta,p\right)\simeq0.8$ for
$1-p=0.1$, which is congruent with the results from Fig.~\ref{fig6: Granular Reliability--10km}.
For $1-p\apprge0.1$, the meta distribution and its Beta estimate
are relatively well matched, predicting $F_{\mathrm{r}}\left(\beta,0.9\right)\in[0.75,0.8]$
for both urban and suburban cases. The match between the simulation
and the Beta estimate break down when $1-p\apprle0.03$. Here, the
Beta distribution over-estimates the fraction of PPPs that lead to
very low conditional outage. For $R=200$\,m (see Fig.~\ref{fig5: MDShort}),
the meta distribution shows a clear bimodal behavior, with a flat
CDF for at least one order of magnitude of conditional outages. This
effect is present for both urban and suburban intersections. The estimated
Beta distribution cannot follow this trend, and thus severely overestimates
$F_{\mathrm{r}}\left(\beta,p\right)$ for small conditional outage
probabilities. 

Altogether, these results indicate that the average performance alone
is not an adequate metric to assess communication reliability, but
with different reasons for large and small road segments comprising
an intersection. In addition, approximations using a Beta distribution
can lead to overestimation in the low conditional outage regime.

\section{\label{Sec7: Conclusion}Conclusion}

We evaluated the average and the fine-grained reliability for interference-limited
V2V communications. The average communication reliability can be determined
in closed-form notation using techniques from stochastic geometry.
However, as the average is taken with respect to the channel environment
and the interfering vehicular traffic, average reliability hides the
underlying attributes of the random traffic. The meta distribution
uncovers this, by describing the standalone reliability incurred from
each PPP. We performed extensive Monte Carlo simulations to estimate
both the average and the fine-grained communications reliability.
Our results indicate that for small road segments, the meta distribution
is bimodal so that PPPs are either not causing any interference or
a lot of interference. Moreover, approximations of the meta distribution
with a Beta distribution tend to be loose for short road segments.

\appendices{}

\section*{Acknowledgments}

This research work is supported, in part, by the EU-H2020 Marie Sk\l odowska-Curie
Individual Fellowship, EU-MARSS-5G project, Grant No. 659933; the
Ericsson Research Foundation, Grant No. FOSTIFT-16:043-17:054; the
VINNOVA COPPLAR project, Grant No. 2015-04849; and the EU-H2020 HIGHTS
project, Grant No. MG-3.5a-2014-636537. The authors are also grateful
to Prof. Martin Haenggi (University of Notre Dame, IN, USA) for discussions
regarding the meta distribution.

\bibliographystyle{IEEEtran}
\bibliography{refJP17}

\end{document}